\documentclass[runningheads]{llncs}
\usepackage[T1]{fontenc}
\usepackage{longtable}
\usepackage{graphicx}
\usepackage{hyperref}
\usepackage{color}
\usepackage{float}
\usepackage[frozencache=true, cachedir=minted-cache]{minted}

\usepackage{alltt}
\usepackage{xspace}
\newcommand{\tool}[1]{\textsc{#1}\xspace}
\usepackage{listings}

\usepackage{tcolorbox}

\tcbset{
  colback=white,        
  colframe=black,       
  coltitle=black,       
  colbacktitle=white,   
  boxrule=0.4pt,        
  arc=1mm,              
  left=1pt,             
  right=1pt,            
  top=1pt,              
  bottom=1pt,           
  fontupper=\normalfont\small, 
  fonttitle=\bfseries\small,   
}

\usepackage{hyperref}

\makeatletter
\newcommand{\printfnsymbol}[1]{%
  \textsuperscript{\@fnsymbol{#1}}%
}
\makeatother

\usepackage{enumitem}
\usepackage{caption}
\begin{document}
\title{Poster: Machine Learning for Vulnerability Detection as Target Oracle in Automated Fuzz Driver Generation}
\titlerunning{ML4VD as Target Oracle in AFDG}
%
\author{Gianpietro Castiglione\thanks{These authors contributed equally}\and
Marcello Maugeri\printfnsymbol{1}\and
Giampaolo Bella}
\authorrunning{Castiglione, Maugeri and Bella}
\institute{University of Catania, Italy \\
\email{\{gianpietro.castiglione,marcello.maugeri\}@phd.unict.it} \\
\email{giampaolo.bella@unict.it}}
\maketitle
\begin{abstract}
In vulnerability detection, machine learning has been used as an effective static analysis technique, although it suffers from a significant rate of false positives.
Contextually, in vulnerability discovery, fuzzing has been used as an effective dynamic analysis technique, although it requires manually writing fuzz drivers.
Fuzz drivers usually target a limited subset of functions in a library that must be chosen according to certain criteria, e.g., the depth of a function, the number of paths.
These criteria are verified by components called \textit{target oracles}.
In this work, we propose an automated fuzz driver generation workflow composed of: (1) identifying a likely vulnerable function by leveraging a machine learning for vulnerability detection model as a target oracle, (2) automatically generating fuzz drivers, (3) fuzzing the target function to find bugs which could confirm the vulnerability inferred by the target oracle.
We show our method on an existing vulnerability in \tool{libgd}, with a plan for large-scale evaluation.

\end{abstract}

\section{Introduction}
In recent years, two automated techniques have emerged for finding 0-day vulnerabilities in functions: \textit{Machine Learning for Vulnerability Detection (ML4VD)} \cite{10.5555/3698900.3699138} and \textit{Fuzzing} for vulnerability discovery.

ML4VD models, when trained on large datasets, are employed to determine whether a given set of functions may exhibit specific vulnerabilities.
However, the method involves static analysis, which cannot verify the vulnerability at runtime, may suffer from a significant false-positive rate~\cite{chan2023diversevul}, and ultimately, top-performing models may not be able to differentiate between vulnerable functions and patched functions~\cite{10.5555/3698900.3699138}.

On the other hand, fuzzing employs dynamic analysis, reducing false positives from static analysis.
However, no push-button fuzzing technique exists, as a \textit{fuzzer} requires a \textit{fuzz driver}, a test harness for parsing inputs and invoking the target function that, in turn, requires deep knowledge about the target function and the corresponding library and extensive manual work~\cite{Babi2019}.
\textit{Automated Fuzz Driver Generation (AFDG)}, despite its challenges, solves the burden, with \tool{OSS-Fuzz-Gen}~\cite{liu2024ossfuzzgen} standing on top due to the use of Large Language Models (LLMs).
Contextually, a library could include several functions, and \textit{target oracles} are employed to identify \textit{interesting} functions~\cite{weissberg2024sok}, i.e. functions determined to be likely interesting targets.
For example, \tool{OSS-Fuzz-Gen} prioritises functions relying on \tool{Fuzz Introspector}'s\footnote{\url{https://github.com/ossf/fuzz-introspector}} heuristics, which mainly consider the cyclomatic complexity of the target function or the simplicity to generate the fuzz driver.
Nevertheless, these heuristics do not account for the likelihood of a function being vulnerable.
Hence, we propose a combined method employing ML4VD models as a target oracle to prioritise relevant functions for the AFDG.

Considering these assumptions, this study is based on the following research questions:
\begin{enumerate}[label=\textbf{RQ\arabic*}, , align=left, leftmargin=!, labelwidth=2em, labelsep=1em]
    \item Can machine learning for vulnerability detection be an effective target oracle for automated fuzz driver generation?
    \item To what extent can such a combined method confirm the true positives, and/or reduce the number of false positives?
\end{enumerate}

In this work, we present the design and workflow that combines the two techniques in Section~\ref{sec:design}.
We validate the method by selecting a confirmed vulnerable function from the \tool{DiverseVul} dataset and successfully applying \tool{OSS-Fuzz-Gen} to generate a fuzz driver that triggers the vulnerability.
The target function originates from a project already included in the \tool{OSS-Fuzz} infrastructure\footnote{\url{https://github.com/google/oss-fuzz}}, ensuring compatibility with \tool{OSS-Fuzz-Gen}, but is not currently covered by an existing fuzz target.
This allows us to generate a novel fuzz driver and achieve previously unreached code coverage.
The complete experimental setup is detailed in Section~\ref{sec:case}.
Subsequently, we examine the state-of-the-art, discussing insights or differences from our technique in Section~\ref{sec:related}.
Finally, we discuss our practical contribution and propose a research plan for an in-depth evaluation in Section~\ref{sec:conclusions}.

\section{Design}
\label{sec:design}
To address the research questions, the proposed design employs two main techniques: a ML4VD model as the vulnerability detection component (static analysis of the target function code), and AFDG, for ultimately applying fuzzing as the discovery component (dynamic analysis that uncovers vulnerabilities during execution of the target function).

\begin{figure}[H]
    \centering
    \includegraphics[width=\linewidth]{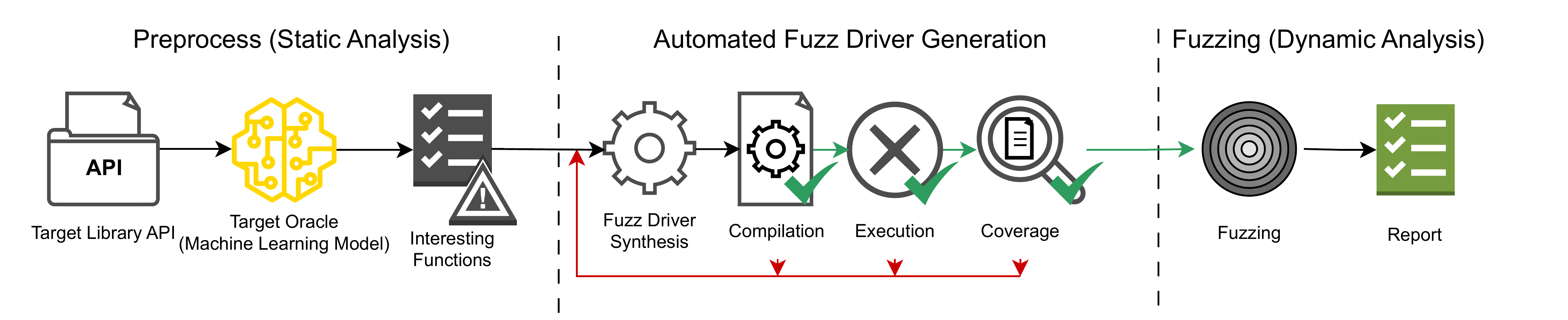}
    \caption{Workflow of the proposed method}
    \label{fig:workflow}
\end{figure}

The overall workflow of the proposed method, illustrated in Figure~\ref{fig:workflow}, comprises three steps.
It represents a generalised pipeline for AFDG and emphasises the main contribution of this work.
Namely, the use of the ML4VD model as a target oracle.

\paragraph{Preprocess (Static Analysis).}
Our method first identifies functions in the target library API that are likely to be vulnerable.
The objective is accomplished by an ML4VD model, which performs a static analysis of each function and flags those that may present one or more weaknesses according to the \textit{CWE} framework by MITRE.
Ultimately, the flagged functions are considered potentially interesting functions for further analysis by fuzzing.
\vspace{-0.2cm}
\paragraph{Automated Fuzz Driver Generation.}
Subsequently, the interesting functions are selected for the AFDG process to begin.
This is an iterative process in which the fuzz driver synthesis produces a candidate fuzz driver. The candidate must (1) compile, (2) execute without immediate logical failure, and (3) gather sufficient coverage to ensure the input is correctly injected.
Once these steps are met, the candidate proceeds to the next stage of the workflow.

\vspace{-0.2cm}
\paragraph{Fuzzing (Dynamic Analysis).}
At this stage, the fuzzing process begins, leveraging the previously generated fuzz driver to inject the target function with a large volume of malformed inputs.
At the expiration of the time budget, the fuzzer either discovers a crashing input that confirms the vulnerability or does not.

\section{Case Study: \tool{libgd} Library} 
\label{sec:case}
\paragraph{Target Selection.}
To evaluate the design of the proposed method, we conducted initial experiments on the \tool{libgd}\footnote{\url{https://github.com/libgd/libgd}} library.
In particular, we assumed to have already a ML4VD model trained on the \tool{DiverseVul} dataset~\cite{chan2023diversevul}, which is considered the best collection of vulnerable functions in C/C++.

Subsequently, we selected among the projects in \tool{DiverseVul} one already included in the \tool{OSS-Fuzz} infrastructure and with \tool{Fuzz Introspector} reports available.
We used such requirements to identify functions currently not covered by existing fuzz drivers in such a project.
Consequently, we directly take an uncovered function labelled with CWEs as most likely classified as vulnerable, belonging to the dataset itself.

\begin{figure}
    \caption{Function Signature of \textit{gdImageWebpPtr}}
    \hspace{-2.8cm}
    \includegraphics[trim=30 575 20 110, clip, width=1.5\linewidth]{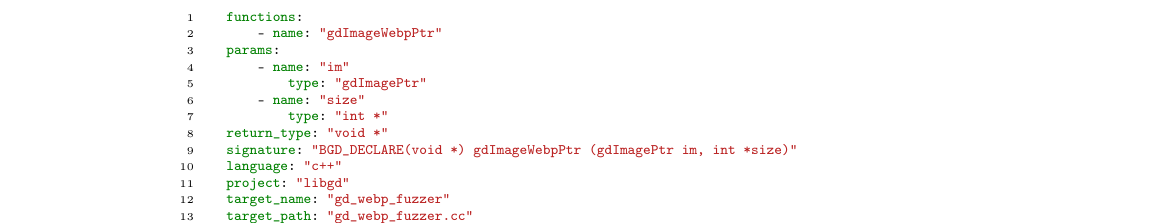}
    \label{fig:sig}
\end{figure}

In particular, we chose the function \textit{gdImageWebpPtr}, labelled as weak to \textbf{CWE-415: Double Free}\footnote{\url{https://cwe.mitre.org/data/definitions/415.html}}, which is confirmed by \textbf{CVE-2016-6912}\footnote{\url{https://nvd.nist.gov/vuln/detail/CVE-2016-6912}}.
\paragraph{Experimental Setting.}
Subsequently, we employed \tool{OSS-Fuzz-Gen} to generate a fuzz driver.
Initially, \tool{OSS-Fuzz-Gen} relies on \tool{Fuzz Introspector} to retrieve the target function signature and its corresponding arguments, which are provided in YAML format, as illustrated in Figure~\ref{fig:sig}.
Since the target function is selected by the target oracle, we assume that our ML4VD model has flagged the \textit{gdImageWebpPtr} function as potentially vulnerable.

After this minimal setup, the AFDG process starts with the creation of a prompt from a template to be provided to a Large Language Model (LLM).
The prompt template is shown in Figure \ref{box:oss}.
In particular, we used the standard prompt template provided in the official repository of \tool{OSS-Fuzz-Gen}; the only difference is the embedded information about the target's weaknesses we added in the prompt, which is shown in red.

\captionof{figure}{OSS-Fuzz-Gen prompt}
\begin{tcolorbox}[title= OSS-Fuzz-Gen Prompt] 
\label{box:oss}
\scriptsize
\begin{tcolorbox}[title= 1 - System Prompt]  
You are a security testing engineer who wants to write a C++ program to discover memory corruption vulnerabilities in a given function-under-test [...]
\end{tcolorbox}
\begin{tcolorbox}[title= 2 - C++ Specific Instructions]  
Use <code>FuzzedDataProvider</code> to generate these inputs [...]
\end{tcolorbox}
\begin{tcolorbox}[title=3 - Instructions and Examples]
[...]
Do not create new variables with the same names as existing variables.
WRONG:
<code> \\
int LLVMFuzzerTestOneInput(const uint8\_t *data, size\_t size) \{ \\
  void* data = Foo(); \\
\} </code> [...]
\end{tcolorbox}
\end{tcolorbox}
\begin{tcolorbox}[title= OSS-Fuzz-Gen Prompt (continue)]
\begin{tcolorbox}[title= 4 - Problem Statement]
Your goal is to write a fuzzing harness for the provided function-under-test signature [...] 
<function signature> \\
\textcolor{red}{Note: The function is a candidate for the vulnerability CWE-415 (Double Free)} \end{tcolorbox}    
\end{tcolorbox}
\normalsize

In our initial experiments, we employed \textit{GPT-4} with a temperature setting of 0.
Although the model successfully generated valid fuzz drivers on first attempts, the fuzz drivers struggled to find a bug and ultimately confirm the vulnerability.
This is because the specific vulnerability can be detected whether \textit{gdImageWebpPtr} is invoked by passing a sufficiently large image.
Instead, the fuzz drivers presented a hard-coded limit cap on the size of the image, likely to prevent out-of-memory errors during the fuzzing campaign.
To solve the issue, we instructed the model to allow large images without setting a cap.
From the initial experiments, we can conclude that if a valid fuzz driver fails to find a bug, this does not rule out a vulnerability, and further instructions based on the expected vulnerability could be needed.
Ultimately, our contribution aims to (1) prioritise functions likely to expose a weakness, (2) confirm a vulnerability whenever a critical bug is found.
\vspace{-0.05cm}

\section{Related Works}
\label{sec:related}
Risse \emph{et al.}~\cite{10.5555/3698900.3699138} have shown that top-performing ML4VD models are unable to distinguish between functions that contain a vulnerability and functions where the vulnerability is patched.
Consequently, without a definitive solution, we expect an increase in false positives over time, which could be mitigated by our method.
The state-of-the-art employs \textit{Directed Greybox Fuzzing (DGF)} to steer the generation of inputs that can reach a specific program location.
Zhu \emph{et al.}~\cite{DBLP:journals/corr/abs-2010-12149}, Yu \emph{et al}.~\cite{Yu2022VulnerabilityorientedDF}, already employed ML4VD as target oracle for DFG.
However, DFG itself does not ensure that the target function can be reached.
For example, a compiled binary could never call a library function.
Our work employs automated fuzz driver generation to generate fuzz drivers which call the target function.

\section{Considerations and Future Works}
\label{sec:conclusions}

\paragraph{Contributions.}
\tool{Fuzz Introspector} uses two heuristics to determine functions likely to be interesting targets.
However, the heuristics fall short of considering interesting only functions having high cyclomatic complexity and containing \textit{parse} in their name (Heuristic 1), and accepting the same argument types as the fuzzing interface \textit{LLVMFuzzerTestOneInput} (Heuristic 2).
Our ultimate aim is to propose a third heuristic, in which an ML4VD model identifies potentially vulnerable functions.

\vspace{-0.2cm}
\paragraph{Threats to Validity.}
As the case study involves a CVE from 2015, this may raise a question about whether the code to reproduce such vulnerability is memorised by the model from the training data.
Future studies will focus on evaluating the effectiveness of novel candidate functions.
\vspace{-0.2cm}
\paragraph{Future Work.}
Future work primarily focuses on applying the implemented method to a broader range of functions.
Our plan encompasses the integration of an ML4VD model as a target oracle (third heuristic) into \tool{Fuzz Introspector} and evaluating its effectiveness on at least ten projects from both \tool{OSS-Fuzz} and \tool{DiverseVul}, relying on \tool{OSS-Fuzz-Gen} for AFDG.

In particular, to answer \textbf{RQ1}, we plan to evaluate our target oracle in novel interesting functions, i.e. not included in the dataset.

To answer \textbf{RQ2}, we will focus on functions in a test set, measuring the precision of both the target oracle and, ultimately, the AFDG process.

\bibliographystyle{splncs04}
\bibliography{bibliography}
\end{document}